\documentstyle[prd,eqsecnum,aps,amsfonts]{revtex}

\begin{document}
\draft
\title{Collapsing regions and black hole formation}
\author{Gregory A. Burnett}
\address{Department of Physics, University of Florida,
Gainesville, Florida\ \ 32611}
\date{4 August 1995}
\twocolumn[
\maketitle
\widetext
\begin{abstract}
Up to a conjecture in Riemannian geometry, we significantly strengthen
a recent theorem of Eardley by proving that a compact region in an
initial data surface that is collapsing sufficiently fast in
comparison to its surface-to-volume ratio must contain a future
trapped region.  In addition to establishing this stronger result, the
geometrical argument used does not require any asymptotic or energy
conditions on the initial data.  It follows that if such a region can
be found in an asymptotically flat Cauchy surface of a spacetime
satisfying the null-convergence condition, the spacetime must contain
a black hole with the future trapped region therein.  Further, up to
another conjecture, we prove a strengthened version of our theorem by
arguing that if a certain function (defined on the collection of
compact subsets of the initial data surface that are themselves
three-dimensional manifolds with boundary) is not strictly positive,
then the initial data surface must contain a future trapped region.
As a byproduct of this work, we offer a slightly generalized notion of
a future trapped region as well as a new proof that future trapped
regions lie within the black hole region.
\end{abstract}
]
\narrowtext

\section{Introduction} \label{sec:introduction}

Given an initial data set for the gravitational field $(\Sigma,
g_{ab}, K_{ab})$ \cite{ids} associated with a Cauchy surface in an
asymptotically flat spacetime, can we tell whether gravitational
collapse has proceeded to such a point that a black hole has formed in
that spacetime?  In principle, the answer is yes: Given a complete
description of the matter fields in the spacetime (i.e., their initial
data and evolution equations), then using Einstein's equation, evolve
the initial data to reconstruct the entire spacetime and then see
whether the spacetime has a nonempty black-hole region and, if so,
whether and where it intersects $\Sigma$.  In practice, however,
carrying out this construction is a highly nontrivial task, even in
the vacuum case and even when done numerically \cite{Anninos94}.

While there is currently no simple algorithm for determining from an
initial data set whether a black hole has formed, if a future trapped
region exists in $\Sigma$, it must lie within the black-hole region,
provided the spacetime satisfies the null-convergence condition
\cite{ncc,Wald84,HawkingEllis73}.  Recall that a closed subset $C$ of
$\Sigma$ having the structure of a three-manifold with smooth (or at
least $C^2$) boundary, bounded away from spatial infinity, is said to
be a future trapped region if the convergence of the future-directed
null geodesics orthogonal to $\partial C$ and outward directed (in the
sense that the projection of null geodesic tangent vectors on $C$ into
$\Sigma$ point outward from $C$) is non-negative everywhere on
$\partial C$ \cite{Wald84}.  Denoting the induced metric on $\partial
C$ by $h_{ab}$, the mean extrinsic curvature \cite{mean} of $\partial
C$ in $\Sigma$ by $H$, and the extrinsic curvature of $\Sigma$ in the
spacetime by $K_{ab}$, $C$ is a future trapped region if
\begin{equation} \label{ftr}
H \le K_{ab}h^{ab}
\end{equation}
on $\partial C$.  Likewise, the total future trapped region of
$\Sigma$ (being the closure of the union of all future trapped regions
in $\Sigma$) along with its boundary ${\cal A}$ (the future apparent
horizon, on which $H = K_{ab}h^{ab}$), is contained within the
black-hole region (under the same conditions as before)
\cite{apparent}.  So, in initial data sets with a non-empty total
future trapped region, gravitational collapse has proceeded
sufficiently far so that black holes have formed.

Are there any simple conditions that guarantee the existence of a
future trapped region in an initial data set?  Thorne's ``hoop
conjecture'' offers a test of this type: If a body of mass $M$ is
sufficiently compact so that a hoop of circumference $4\pi M$ can
encircle the body no matter how it is rotated there about, then the
body must be contained within a horizon \cite{Thorne72,Flanagan91}.
While a precise version of this conjecture remains to be proven,
Schoen and Yau have proven that an initial data set containing a
region $\Omega$ with sufficient matter density must contain a future
or past apparent horizon \cite{SchoenYau83}.  This interesting result
has the slight weaknesses that its requirement on the matter content
is so strict that arbitrarily small vacuum regions in $\Omega$ are not
allowed and that, as a time-symmetric theorem, we cannot conclude that
the apparent horizon must be a future horizon.  (Of course, this last
criticism can be avoided by restricting oneself to initial data
sets that do not contain past trapped regions.)\ \ Further, a number
of sufficient conditions have been found for spherically symmetric
initial data sets \cite{ss}.  More recently, using Jang's equation and
its properties as established by Schoen and Yau \cite{SchoenYau81},
Eardley has recently provided a remarkably simple proof of the
following theorem \cite{Eardley95,Eardley93}.

{\it Theorem (Eardley)}.  Fix an asymptotically flat initial data set
for the gravitational field $(\Sigma, g_{ab}, K_{ab})$ \cite{ids}
satisfying the dominant-energy condition.  If there exists a compact
region $\Omega \subset \Sigma$ such that $K_{ab}(g^{ab} - n^a n^b)$ is
no less than the surface-to-volume ratio of $\Omega$ for all unit
vectors $n^a$ everywhere on $\Omega$, then $\Sigma$ must contain an
apparent horizon.

Were the apparent horizon a {\em future} apparent horizon, then, as
noted, the conditions of this theorem would guarantee that
gravitational collapse has proceeded sufficiently far that a
black-hole region has formed.  Unfortunately, as given, the theorem
alone does not allow one to draw this conclusion as it suffers from
the same problem of Schoen and Yau's theorem \cite{SchoenYau83} in
that the possibility that that all such horizons will be past apparent
horizons has not been eliminated.  However, the time-asymmetry in the
hypotheses of Eardley's theorem ($K^a{}_a$ is strictly positive on
$\Omega$ indicating that the region is collapsing ``on average'') is a
strong indication that there should be a future apparent horizon
somewhere in $\Sigma$.

By changing our viewpoint, Eardley's theorem suggests an alternative
argument having the advantage of producing a strengthened version of
the above theorem under weaker hypotheses.  In particular, we can now
show that $\Sigma$ must in fact contain a future apparent horizon.
Further, this argument has an entirely geometric character, which is
to be compared to Eardley's argument, which, through the use of Jang's
equation, has an analytic character.

To begin, notice that the induced metric $h^{ab}$ on a two-surface
${\cal S} \subset \Sigma$ can be written as $h^{ab} = g^{ab} - n^a
n^b$, where $n^a$ is either of the two unit-normal vectors to ${\cal
S}$.  Therefore, the hypothesis of the above theorem guarantees that,
on ${\cal S}$, $K_{ab}h^{ab}$ is bounded from below by the
surface-to-volume ratio of $\Omega$, which we denote by
$\sigma(\Omega)$.  Notice that this bound is independent of the
two-surface in $\Omega$.  This suggests that if there exists a region
in $\Omega$ (having the structure of a three-manifold with boundary)
whose boundary's mean extrinsic curvature $H$ is bounded above by
$\sigma(\Omega)$, then Eq.~(\ref{ftr}) would hold on the boundary, and
hence the region would be a future trapped region.

Does such a region exist in $\Omega$?  The appearance of the
surface-to-volume ratio $\sigma$ in Eardley's theorem suggests that we
study this quantity as function on the collection of regions $C$ in
$\Omega$.  Consider a region $C \subset \Omega$ that is ``nearly
degenerate'' in the sense that it is either flat like a pancake of
thickness $r$, thin like a cigar of radius $r$, or small like a sphere
of radius $r$.  Then, we expect (as in the flat space case) that
$\sigma(C) \approx \text{const}/r$, for $r$ sufficiently small,
showing that regions that are nearly degenerate in the sense that they
are small in one or more dimensions have very large surface-to-volume
ratios.  This suggests that there is some sufficiently well-behaved
region in $\Omega$ having minimal surface-to-volume ratio.  In fact,
we conjecture that there always exists a region $\hat{C} \subset
\Omega$, having the structure of a differentiable manifold with
boundary, that minimizes $\sigma$ over such regions $C \subset
\Omega$.  (See conjecture~1 in Sec.~\ref{sec:conjecture}.)\ \
Remarkably, it then follows that $H \le \sigma(\Omega)$ on any open
subset of $\partial\hat{C}$ where the surface is $C^2$, i.e., on the
portion having a well-defined and continuous extrinsic curvature.
(The proof of this fact is given in Sec.~\ref{sec:lemma1}.)\ \

Putting this all together, we have
\begin{equation} \label{proof1}
H \le \sigma(\Omega) \le K_{ab}h^{ab}
\end{equation}
on the open subset of $\partial\hat{C}$ that is $C^2$, where the first
inequality is a consequence of the minimizing property of $\hat{C}$
and the second follows by hypothesis.  Were $\partial\hat{C}$
everywhere $C^2$, then $\hat{C}$ would be a future trapped region.  As
explained in Sec.~\ref{sec:conjecture}, this is not always the case.
However, it is expected that $\partial\hat{C}$ is sufficiently
well-behaved so that Eq.~(\ref{proof1}) holds over a sufficiently
large subset of $\partial\hat{C}$ that $\hat{C}$ is indeed a future
trapped region, in the sense that it must lie in the black-hole region
of the spacetime.  In particular, we conjecture that $\partial\hat{C}$
is everywhere $C^{2-}$ (see below) and $C^2$ everywhere except on a
closed set ${\cal Z}$ of measure zero.  (See Sec.~\ref{sec:conjecture}
for the statement and discussion of this conjecture.)\ \ Therefore,
although Eq.~(\ref{ftr}) may not hold everywhere on $\partial\hat{C}$,
it does hold on $\partial\hat{C} \setminus {\cal Z}$, which, with the
fact that the surface is $C^{2-}$, is sufficient to guarantee the the
region is trapped.  (See theorem~5 in Sec.~\ref{sec:trapped_notc2}.)\
\ This proves the following strengthened version of Eardley's theorem.

{\it Theorem 1.}  Fix an initial data set for the gravitational field
$(\Sigma,g_{ab},K_{ab})$ \cite{ids} and fix a subset $\Omega \subset
\Sigma$ that is a compact three-manifold with $C^2$ boundary.  If
$K_{ab}h^{ab}$ is no less than the surface-to-volume ratio of $\Omega$
for all rank~2 orthogonal projection maps $h^a{}_b$ \cite{projop}
everywhere on $\Omega$, then there exists a future trapped region in
$\Omega$, provided conjecture~1 (stated in Sec.~\ref{sec:conjecture})
holds for $(\Omega,g_{ab})$.

Denoting the eigenvalues of $K^a{}_b$ by $(k_1,k_2,k_3)$, ordered so
that $k_1 \le k_2 \le k_3$, it is worth noting that the minimum of
$K_{ab}h^{ab}$ over all rank~2 orthogonal projection maps $h^a{}_b$
\cite{projop} is precisely $k_1 + k_2$.  Therefore, the sole condition
of theorem~1 is that the sum of the two lesser principal (extrinsic)
curvatures be no less than the surface-to-volume ratio of $\Omega$,
everywhere on $\Omega$.

We can now assert (assuming conjecture~1) that if such a region
$\Omega$ exists in a Cauchy surface of an asymptotically flat
spacetime \cite{Wald84} satisfying the null-convergence condition
\cite{ncc}, the spacetime must contain a black hole with the future
trapped region therein.

Comparing the two theorems, we see that while neither locates the
future apparent horizon, theorem~1 does tell us some subset of
$\Omega$, namely $\hat{C}$, is contained within the future apparent
horizon.  Further, we see that theorem~1 dispenses with the asymptotic
and energy conditions that were needed by Eardley because of their use
in Schoen and Yau's analysis of Jang's equation.

The remainder of this work is organized as follows.  In
Sec.~\ref{sec:lemma1}, we prove the lemma providing the bound $H \le
\sigma(\Omega)$ on $\partial\hat{C}$.  In Sec.~\ref{sec:general}, we
review Eardley's argument and then present a strengthened version of
theorem~1.  In Sec.~\ref{sec:conjecture}, we state and discuss the two
conjectures in Riemannian geometry needed for this work.  In
Sec.~\ref{sec:trapped}, we offer a new proof that future trapped
regions are trapped, which is then modified to establish the same
result for our weaker notion a future trapped region, and then we
discuss the possibility of further extending the notion of a future
trapped region.  Lastly, in Sec.~\ref{sec:discussion}, we discuss the
strengths and weakness of our results.

Our conventions are those of Ref.~\cite{Wald84} with the notable
exception that our sign convention for the extrinsic curvature of our
initial data surfaces is such that positive $K$ is associated with
collapse in the sense that it measures the {\em convergence} of
future-directed geodesic normals to the surface.  On the other hand,
$H$ measures the {\em divergence} of the outward geodesic normals to
the surface of a region within an initial data surface.

Recall that a map between manifolds is said to be $C^{k-}$ if the the
mapping is $C^{k-1}$ and its $(k-1)$-order derivatives of the
functions defining the mapping are locally Lipschitz
\cite{HawkingEllis73,Choquet-Bruhat89}.  Thus, a $C^{2-}$ embedded
surface is $C^1$ and the derivative of the embedding map is locally
Lipschitz.

It proves very convenient to make the following definitions.  Given a
manifold $N$ (possibly with boundary), define ${\cal C}^k(N)$
[${\cal C}^{k-}(N)$] to be the collection of compact subsets of $N$
having the structure of a manifold with $C^k$ ($C^{k-}$) boundary.
It is useful to keep in mind that
\begin{equation}
{\cal C}^0(N) \supset {\cal C}^{1-}(N) \supset {\cal C}^1(N) \supset
{\cal C}^{2-}(N) \supset {\cal C}^2(N) \supset \cdots.
\end{equation}
Elements of ${\cal C}^k(N)$ and ${\cal C}^{k-}(N)$ need not be
connected, i.e., they can have many connected components.  Further, if
$\Omega \in {\cal C}^k(N)$, then $\Omega \in {\cal C}^k(\Omega)$, and,
similarly, if $\Omega \in {\cal C}^{k-}(N)$, then $\Omega \in {\cal
C}^{k-}(\Omega)$.

Lastly, for a map $\phi: A \rightarrow B$, $\phi[A]$ denotes the image
of $A$ in $B$, $A \setminus B$ denotes the set of elements in $A$ that
are not in $B$, and $\overline{A}$ denotes the closure of $A$.

\section{Proof that $H \le \sigma(\Omega)$ on $\partial\hat{C}$}
\label{sec:lemma1}

Denote the surface area, volume, and surface-to-volume ratio of a
region $C \in {\cal C}^1(\Sigma)$ by $A(C)$, $V(C)$, and $\sigma(C) =
A(C)/V(C)$, respectively.  More explicitly,
\begin{mathletters} \label{def_AV}
\begin{eqnarray}
A(C) & = & \int_{\partial C} \epsilon_{ab}, \\
V(C) & = & \int_{C} \epsilon_{abc},
\end{eqnarray}
\end{mathletters}
where $\epsilon_{abc}$ is the volume element constructed from $g_{ab}$
and $\epsilon_{ab}$ is the volume element constructed from the metric
$h_{ab}$ induced on $\partial C$ by $g_{ab}$.

{\it Lemma 1.}  Fix a pair $(\Omega,g_{ab})$, where $\Omega$ is a
compact three-dimensional manifold with $C^1$ boundary and $g_{ab}$ is
a smooth Riemannian metric on $\Omega$.  If $\hat{C} \in {\cal
C}^1(\Omega)$ is such that $\sigma(\hat{C}) \le \sigma(C)$ for all $C
\in {\cal C}^1(\Omega)$ and $O$ is an open subset of $\partial\hat{C}$
where the surface is $C^2$, then $H \le \sigma(\hat{C}) \le
\sigma(\Omega)$ on $O$, where $H$ is the mean extrinsic curvature of
$\partial\hat{C}$ \cite{mean}.  If, further, $O$ is in the interior of
$\Omega$, then $H = \sigma(\hat{C})$ on $O$.

{\it Proof.}  The idea of the proof is simple: We calculate $\sigma$
as a function along certain well-behaved curves in ${\cal
C}^1(\Omega)$ containing $\hat{C}$, calculate its derivative at
$\hat{C}$, and then use the fact that $\hat{C}$ minimizes $\sigma$ in
${\cal C}^1(\Omega)$.

Although there are many curves in ${\cal C}^1(\Omega)$, by which we
mean one-parameter family of regions $C_\lambda \in {\cal
C}^1(\Omega)$, for simplicity, we shall restrict ourselves to families
arising from a smooth deformation of a region $C \in {\cal C}^1(\Omega)$
in the sense that $C_\lambda = \phi_\lambda[C]$ for some one-parameter
family of maps $\phi_\lambda: \Omega \rightarrow \Omega$ such that
$\phi_\lambda$ is a diffeomorphism between $\Omega$ and $\phi_\lambda[
\Omega]$, with $\phi_0$ being the identity map on $\Omega$.  Our
requirement that $\Omega$ and $\phi_\lambda[\Omega]$ be diffeomorphic
is sufficient to guarantee that $\partial(\phi_\lambda[C]) =
\phi_\lambda[\partial C]$, which makes the following calculations
easier than they would be otherwise.

A particularly simple class of such deformations, which is sufficient
for our purposes, are those associated with the flows of fixed vector
fields on $\Omega$ \cite{Choquet-Bruhat89}.  [That is, given a fixed
vector field $\xi^a$, for $p \in \Omega$, $\phi_\lambda(p)$ is the
point along the integral curve of $\xi^a$ containing $p$ a parameter
distance $\lambda$ from $p$.]\ \ In order that these deformations be
well defined on all of $\Omega$ for some positive $\lambda$, it is
necessary to restrict ourselves to vector fields that are inward
pointing everywhere on $\partial\Omega$ (where we consider vectors
tangent to $\partial\Omega$ as inward pointing, so $\xi^k n_k \le 0$
everywhere on $\partial\Omega$, where $n^k$ is the unit outward normal
to $\partial\Omega$).  Otherwise, a point $p \in \partial\Omega$ where
$\xi^a$ is strictly outward pointing would be mapped ``out of''
$\Omega$, and hence the deformation constructed from it would not be
defined for any positive $\lambda$, no matter how small.  A
deformation $\phi_\lambda$ constructed from an inward pointing vector
field is well defined for all $\lambda \ge 0$ and is a diffeomorphism
between $\Omega$ and $\phi_\lambda[\Omega]$.

Fix any inward pointing vector field $\xi^a$ whose support intersects
$\partial C$ within $O$ and construct its one-parameter family of
deformations $\phi_\lambda$.  Evaluating $A$ and $V$ on $C_\lambda =
\phi_\lambda[C]$ using Eqs.~(\ref{def_AV}), differentiating with
respect to $\lambda$, and then evaluating at $\lambda = 0$, we find
that
\begin{mathletters} \label{perturb}
\begin{eqnarray}
A'(C) & = & \int_{O} H (\xi^k n_k) \epsilon_{ab}, \\
V'(C) & = & \int_{O}   (\xi^k n_k) \epsilon_{ab},
\end{eqnarray}
\end{mathletters}
where $H$ is the mean extrinsic curvature of $\partial C$ and $n^k$ is
the outward unit normal to $\partial C$.  Differentiating the equality
$\sigma(C_\lambda) = A(C_\lambda)/V(C_\lambda)$, evaluating at
$\lambda = 0$, and using Eqs.~(\ref{perturb}), we find that
\begin{equation} \label{perturb_sigma}
\sigma'(C) = {1 \over V(C)} \int_{O}
\biglb(H-\sigma(C)\bigrb) (\xi^k n_k) \epsilon_{ab}.
\end{equation}
Using this equation, we now establish our bound on $H$.

We begin with the case where $O$ is in the interior of $\Omega$.  Fix
any point $p \in O$.  To show that $H(p) = \sigma(\hat{C})$, suppose,
for contradiction, that $H(p) > \sigma(\hat{C})$.  Then, using the
facts that $p$ is in the interior of $\Omega$ and $H$ is continuous at
$p$, it is not difficult to show that there exists an open
neighborhood $N$ of $p$ and a vector field $\xi^a$ such that: (1) the
support of $\xi^a$ is $\overline{N}$; (2) $\partial\hat{C}$ is $C^2$
on $\overline{N} \cap \partial\hat{C}$; (3) $(\overline{N} \cap
\partial\Omega) = \emptyset$; (4) $\biglb(H-\sigma(\hat{C})\bigrb) >
0$ on $N \cap \partial\hat{C}$; (5) $(\xi^k n_k) < 0$ on $N \cap
\partial\hat{C}$.  Notice that $\xi^a$, being zero on
$\partial\Omega$, is inward pointing, so the one-parameter family of
deformations $\phi_\lambda$ constructed from $\xi^a$ is defined for
all $\lambda \ge 0$.  Using Eq.~(\ref{perturb_sigma}), we see that
$\sigma'(\hat{C}) < 0$, which is impossible as otherwise, for
sufficiently small $\lambda$, the region $\phi_\lambda[\hat{C}]$ would
have a smaller surface-to-volume ratio than $\hat{C}$.  Similarly, if
$H(p) < \sigma(\hat{C})$, there exists an open neighborhood $N$ of $p$
and a vector field $\xi^a$ satisfying the above except with the
inequalities in (4) and (5) both reversed.  Using
Eq.~(\ref{perturb_sigma}), we again find that $\sigma'(\hat{C}) < 0$,
which is again a contradiction.  Therefore, $H(p) = \sigma(\hat{C})$,
as claimed.

Otherwise, fix any point $p \in O$.  To show that $H(p) \le
\sigma(\hat{C})$, suppose, for contradiction, that $H(p) >
\sigma(\hat{C})$.  Then, there exists an open neighborhood $N$ of $p$
and a vector field $\xi^a$ satisfying the above with (3) replaced by:
(3') $\xi^a$ is inward pointing on $\overline{N} \cap \partial\Omega$.
Again the one-parameter family of deformations $\phi_\lambda$
constructed from $\xi^a$ is defined for all $\lambda \ge 0$.  Using
Eq.~(\ref{perturb_sigma}), we see that $\sigma'(\hat{C}) < 0$, which
is contradicts the minimality of $\sigma$ at $\hat{C}$.  Therefore,
$H(p) \le \sigma(\hat{C})$, as claimed.

Lastly, that $\sigma(\hat{C}) \le \sigma(\Omega)$ follows simply from
the facts that $\hat{C}$ minimizes $\sigma$ over ${\cal C}^1(\Omega)$
and $\Omega \in {\cal C}^1(\Omega)$.  This completes the proof of
lemma~1.~$\Box$

\section{Strengthening theorem~1}
\label{sec:general}

We begin with two definitions.  First define the scalar field $\kappa$
on $\Sigma$ by setting
\begin{equation} \label{def_kappa}
\kappa(p) = \min_{h^a{}_b} (K_{ab}h^{ab}),
\end{equation}
for each $p \in \Sigma$, where the minimum is over the set of all
rank~2 orthogonal projection maps $h^a{}_b$ at $p$ \cite{projop}.
That is, $\kappa(p)$ is the sum of the two lesser principal
(extrinsic) curvatures at $p$.  Second, for any continuous function
$f$ on $\Sigma$, define the function $W_f$ on ${\cal C}^1(\Sigma)$ by
setting
\begin{equation} \label{def_W}
W_f(C) = \int_C f \epsilon_{abc},
\end{equation}
for each $C \in {\cal C}^1(\Sigma)$.  Note that with $f=1$, $W_1(C) =
V(C)$.

The idea behind the proof of theorem~1 was to use the properties of
the region that minimizes the surface-to-volume ratio $\sigma$ on
${\cal C}^1(\Omega)$.  Noting that $\sigma = A/V = A/W_1$, one way to
proceed in generalizing theorem~1 is to analyze the properties of the
region that minimizes $A/W_\kappa$ on ${\cal C}^1(\Omega)$.  Assuming
the relevant generalized version of conjecture~1 holds, it can be
shown that there is a future trapped region in $\Omega$ provided that
$A(\Omega)/W_\kappa(\Omega) \le 1$ and $\kappa$ is non-negative (and
not everywhere zero) on $\Omega$.  However, such an argument must fail
if $\kappa$ is negative somewhere on $\Omega$ since, by choosing
regions with large area in regions where $\kappa$ is negative, we can
find $C \in {\cal C}^1(\Sigma)$ for which the ratio $A(C)/W_\kappa(C)$
is negative and as large as we wish (i.e., $A/W_\kappa$ has no finite
lower bound in this case).  The fact that $\kappa$ cannot be even
slightly negative on small subsets of $\Omega$ makes this route
unattractive, so we take an alternative path suggested by the argument
Eardley used in proving his theorem.

\subsection{Eardley's argument}

Fix an asymptotically flat initial data set for the gravitational
field $(\Sigma, g_{ab}, K_{ab})$ \cite{ids} with sources satisfying
the dominant energy condition.  Schoen and Yau have shown that such an
initial data set does not contain an apparent horizon (either future
or past) if and only if there exists a scalar field $f$ satisfying
Jang's equation everywhere on $\Sigma$ \cite{SchoenYau81}.  Eardley's
argument is that certain initial data are inheritly incompatible with
the existence of a global solution of Jang's equation, and therefore a
(future or past) apparent horizon must be present.  This argument goes
as follows.

Defining
\begin{equation} \label{defh}
h^a = {D^a f \over \sqrt{1 + D^m f D_m f}},
\end{equation}
where $D_a$ is the derivative operator on $\Sigma$ associated with
$g_{ab}$, Jang's equation takes the simple form
\begin{equation} \label{Jang}
D_a h^a = K_{ab}(g^{ab} - h^a h^b).
\end{equation}
Noting that $h^m h_m < 1$ everywhere, define the scalar field
$\tilde\kappa$ on $\Sigma$ by setting
\begin{equation} \label{def_kappat}
\tilde\kappa(p) = \inf_{|x|<1}\left( K_{ab}(g^{ab}-x^ax^b) \right),
\end{equation}
at each point $p \in \Sigma$, where the infimum is over all vectors
$x^a$ at $p$ with $x^m x_m < 1$.  (By continuity, the value of
$\tilde\kappa$ is unchanged if we modify its definition by taking the
minimum over all vectors $x^a$ at $p$ with $x^m x_m \le 1$.)\ \ Using
the fact that for $x^a \neq 0$
\begin{eqnarray}
K_{ab}(g^{ab}&-&x^ax^b) \nonumber \\
& = & (1-x^mx_m)K_{ab}g^{ab} \nonumber \\
&& + (x^mx_m)K_{ab} \left( g^{ab} - x^a x^b /(x^mx_m) \right),
\end{eqnarray}
it is not difficult that show that
\begin{equation}
\tilde\kappa = \min(K^a{}_a, \kappa) \le \kappa
\end{equation}
at each point.  Define the function $\tilde S$ on ${\cal C}^1(\Sigma)$
by setting
\begin{equation}
\tilde S(C) = A(C) - W_{\tilde\kappa}(C),
\end{equation}
for each $C \in {\cal C}^1(\Sigma)$.

If a global solution of Jang's equation exists, it follows that
\begin{equation} \label{Jang2}
D_a h^a \ge \tilde\kappa,
\end{equation}
everywhere on $\Sigma$.  Integrating Eq.~(\ref{Jang2}) over any region
$C \in {\cal C}^1(\Sigma)$ and using the fact that $h^k n_k < 1$
everywhere on $\partial C$, where $n^k$ is the outward unit normal to
$\partial C$, we find that
\begin{equation}
\tilde S (C) > 0.
\end{equation}
That is, $\tilde S$ is a strictly positive function on ${\cal
C}^1(\Sigma)$.  Therefore, if there exists a region $\Omega \in {\cal
C}^1(\Sigma)$ with $\tilde S (\Omega) \le 0$, a global solution of
Jang's equation cannot exist, and, thus, by Schoen and~Yau's results,
a (future or past) apparent horizon must be present within $\Sigma$.

In this argument, we see that the function $\tilde S$ on ${\cal
C}^1(\Sigma)$ arises rather naturally.  This suggests that we should
attempt to strengthen the above result by showing that when there
exists $\Omega \in {\cal C}^1(\Sigma)$ with $\tilde S (\Omega) \le 0$
a future trapped region must exist within $\Omega$ (without the need
for any asymptotic or stress-energy conditions).  However, it turns
out that we can do a little better using the function $S$ on ${\cal
C}^1(\Sigma)$, defined by
\begin{equation} \label{def_S}
S(C) = A(C) - W_\kappa(C),
\end{equation}
for each $C \in {\cal C}^1(\Sigma)$, rather than $\tilde S$.  Since
$\tilde\kappa \le \kappa$, it follows that $S(C) \le \tilde S(C)$,
and, hence, if $\tilde S(\Omega) \le 0$, then $S(\Omega) \le 0$.  So a
future trapped region theorem using $\tilde S$ follows from such a
theorem for $S$.  (See theorem~3, below).

It is worth noting that for many initial data sets, $S$ and $\tilde S$
will coincide.  Using the facts that $K^a{}_a = k_1 + k_2 + k_3$ and
$\kappa = k_1 + k_2$, it follows that $\tilde\kappa < \kappa$ if and
only if $K_{ab}$ is negative definite ($K_{ab}x^a x^b < 0$ for all
nonzero $x^a$), and $\tilde\kappa = \kappa$ otherwise.  Therefore, if
$K_{ab}$ is nowhere negative definite on $\Sigma$, i.e., nowhere is
the surface positively contracting in all directions, then $S(C) =
\tilde S(C)$ for all $C \in {\cal C}^1(\Sigma)$.

\subsection{New argument}

Our first notable property of $S$ is that any region $\hat{C} \in
{\cal C}^2(\Sigma)$ that is a stationary point of $S$ is a future
trapped region.

{\it Theorem 2.}  If $\hat{C} \in {\cal C}^2(\Sigma)$ is a stationary
point of $S$ (in the sense that $S'(\hat{C}) = 0$ for all smooth
variations of $C$), then $\hat{C}$ is a future trapped region.

{\it Proof.}  Fix any region $C \in {\cal C}^2(\Sigma)$ and any open
subset $O$ of $\partial C$.  Then, for all smooth vector fields
$\xi^a$ whose support intersects $\partial C$ within $O$
\begin{equation} \label{perturb_S}
S'(C) = \int_{O} (H-\kappa) (\xi^k n_k) \epsilon_{ab}.
\end{equation}
Therefore, repeating the argument used in lemma~1 and using the fact
that $\hat{C}$ is in the interior of $\Sigma$ (as $\Sigma$ has no
boundary), we find that $H = \kappa$ on $\partial\hat{C}$.  However,
as $\kappa \le K_{ab}h^{ab}$ on $\partial\hat{C}$, where $h^{ab}$ is
the metric induced on $\partial\hat{C}$, $H \le K_{ab}h^{ab}$ on
$\partial\hat{C}$.  Therefore, $\hat{C}$ is a future trapped
region.~$\Box$

Note that if $\hat{C} \in {\cal C}^2(\Sigma)$ is a local minimum of
$S$ in the sense that there is an open set $N \subset \Sigma$ such
that $S(\hat{C}) \le S(C)$ for all $C \in {\cal C}^2(\Sigma)$ with $C
\subset N$, then $\hat{C}$ is a stationary point of $S$.  Further, for
momentarily static initial data sets ($K_{ab} = 0$ on $\Sigma$),
$\kappa = 0$ on $\Sigma$, so $S$ is simply the surface area of
$C$. Therefore, in this case, the problem of finding stationary points
of $S$ is exactly the problem of finding surfaces whose area is
stationary (in the sense of theorem~2), e.g., minimal two-surfaces
\cite{Osserman86}.

Since finding stationary points of $S$ is a difficult task, it is
desirable to have an alternate condition that guarantees the existence
of a future trapped region.  Mimicking the proof of theorem~1, we fix
a region $\Omega \in {\cal C}^2(\Sigma)$ and then analyze the
properties of a region that minimizes $S$ on ${\cal C}^1(\Omega)$.  If
$S(\Omega) > 0$, $S$ may not have a minimum on ${\cal C}^1(\Omega)$.
To see this, note that as there exist regions with arbitrarily small
surface areas and volumes, $\inf_{{\cal C}^1(\Omega)}(S) \le 0$.  Yet,
for an initial data set with $\kappa \le 0$ (as is the case for a
maximal hypersurface), there is no region $\hat{C} \in {\cal
C}(\Omega)$ that attains the infimum (being zero) as any such region
necessary has $S(\hat{C}) \ge A(\hat{C}) > 0$.  However, for any
region $\Omega$ with $S(\Omega) \le 0$, we conjecture that $S$ does
have a minimum on ${\cal C}^1(\Omega)$.  (In fact, we conjecture that
the minimizing region $\hat{C}$ is a member of ${\cal C}^{2-}(\Omega)$
and has further nice differentiable properties.  See conjecture~2 in
Sec.~\ref{sec:conjecture}.)\ \ The idea behind this conjecture is that
if $\inf_{{\cal C}^1(\Omega)}(S) < 0$ (which is guaranteed to be the
case if $S(\Omega) < 0$), a sequence of regions $C_i$ with $S(C_i)$
approaching this infimum cannot become degenerate in the sense that
their volumes go to zero or their areas become infinite, while if
$\inf_{{\cal C}^1(\Omega)}(S) = 0$, then $S(\Omega) = 0$, so $\Omega$
itself is a minimizing region.  Note that $\inf_{{\cal
C}^1(\Omega)}(S)$ must be finite as
\begin{equation}
\inf_{{\cal C}^1(\Omega)}(S) \ge -\max_\Omega(\kappa)V(\Omega);
\end{equation}
a lower bound that holds even if $\kappa$ is negative somewhere on
$\Omega$.  This is to be compared to the difficulty in establishing a
similar result for the surface-to-volume ratio function $\sigma$ and
lack of any finite lower bound on $A/W_\kappa$ when $\kappa$ is
negative somewhere on $\Omega$.  Using these ideas, the following
theorem shows that if $S$ is not strictly positive on ${\cal
C}^2(\Sigma)$, then $\Sigma$ must contain a future trapped region.

{\it Theorem 3.}  If $S(\Omega) \le 0$ for some $\Omega \in {\cal
C}^2(\Sigma)$, then there exists a future trapped region in $\Omega$,
provided conjecture~2 (stated in Sec.~\ref{sec:conjecture}) holds for
$(\Omega,g_{ab})$.

{\it Proof.}  By conjecture~2, there exists $\hat{C} \in {\cal
C}^{2-}(\Omega)$ that minimizes $S$ on ${\cal C}^1(\Omega)$ and
further $\partial\hat{C}$ is $C^2$ on $\partial\hat{C} \setminus {\cal
Z}$, where ${\cal Z}$ is a closed set of measure zero.  Therefore, for
all one-parameter family of deformations constructed from an inward
pointing vector field on $\Omega$ whose support intersects
$\partial\hat{C}$ where the surface is $C^2$, we have $0 \le
S'(\hat{C})$.  Using Eq.~(\ref{perturb_S}), with $C = \hat{C}$ and
repeating the argument used in lemma~1, we find that $H \le \kappa$ on
$\partial\hat{C} \setminus {\cal Z}$.  However, as $\kappa \le
K_{ab}h^{ab}$ on all of $\partial\hat{C}$, where $h^{ab}$ is the
metric induced on $\partial\hat{C}$, $H \le K_{ab}h^{ab}$ on
$\partial\hat{C} \setminus {\cal Z}$.  Therefore, $\hat{C}$ is a
future trapped region.~$\Box$

Note that if $\kappa \le 0$ on $\Sigma$ (as is the case for maximal
hypersurfaces), then there is no region $\Omega$ meeting the condition
of theorem~3 as $S(\Omega) \ge A(\Omega) > 0$.  Further, the condition
of Eardley's theorem and theorem~1 that $\sigma(\Omega) \le
\min_{\Omega}(\kappa)$ implies that $A(\Omega) \le
\min_{\Omega}(\kappa) V(\Omega) \le W_\kappa(\Omega)$ and, therefore,
$S(\Omega) \le 0$, which is the sole condition of theorem~3.
Therefore, theorem~3 is stronger than theorem~1, which is stronger
than Eardley's theorem.

It is interesting to note that $S(C)$ can be expressed as a pure
surface integral by introducing any vector field $\zeta^a$ on $\Sigma$
(or merely on $\Omega$) having the property that $D_a \zeta^a =
\kappa$, where $D_a$ is the derivative operator associated with the
metric $g_{ab}$.  With this, we have
\begin{equation}
S(C) = \int_{\partial C} (1-\zeta^k n_k) \epsilon_{ab}.
\end{equation}
For instance, a particularly simple choice of $\zeta^a$ is that given
by taking $\zeta^a = D^a \phi$ for a scalar field $\phi$.  Then,
$\phi$ must be a solution of Poisson's equation $D_a D^a \phi =
\kappa$ and can be fixed uniquely by fixing boundary data for $\phi$
on $\partial \Omega$ (e.g., $\phi=0$ on $\partial\Omega$) or a
boundary condition on $\phi$ at infinity (though whether this can
always be accomplished is more subtle).  We will not pursue this
formulation any further here as nothing new seems to gained from this
viewpoint.

In theorems~1 and~3, we have restricted ourselves regions $\Omega$
with $C^2$ boundary for the sake of simplicity, and we expect that
both theorems hold under weaker conditions.  It would seem that the
weakest differentiability condition that should be imposed is that for
which it makes sense for a region to speak of a region being future
trapped.

\section{Two Geometrical Conjectures}
\label{sec:conjecture}

The relevance of theorems~1 and~3 rests heavily upon the following two
conjectures, which we believe to be true.

{\it Conjecture 1.}  Fix a pair $(\Omega,g_{ab})$, where $\Omega$ is a
compact three-dimensional manifold with $C^2$ boundary and $g_{ab}$ is
a smooth Riemannian metric on $\Omega$.  There exists $\hat{C} \in
{\cal C}^{2-}(\Omega)$ such that $\sigma(\hat{C}) \le \sigma(C)$ for
all $C \in {\cal C}^1(\Omega)$.  [In other words, $\sigma$ has a
minimum on ${\cal C}^1(\Omega)$ and a minimizing region is a member of
${\cal C}^{2-}(\Omega)$.]\ \ Further, $\partial\hat{C}$ is $C^2$
everywhere except on the closed set of measure zero given by
$\partial {\cal W}$, where ${\cal W} = (\partial\Omega \cap
\partial\hat{C})$ (and $\partial {\cal W}$ is constructed viewing
${\cal W}$ as a subset of either $\partial\Omega$ or
$\partial\hat{C}$).

{\it Conjecture 2.}  Fix a triple $(\Omega,g_{ab},\kappa)$, where
$\Omega$ is a compact three-dimensional manifold with $C^2$ boundary,
$g_{ab}$ is a smooth Riemannian metric on $\Omega$, and $\kappa$ is a
smooth scalar field on $\Omega$.  If $S(\Omega) \le 0$, then there
exists $\hat{C} \in {\cal C}^{2-}(\Omega)$ such that $S(\hat{C}) \le
S(C)$ for all $C \in {\cal C}^1(\Omega)$.  [In other words, $S$ has a
minimum on ${\cal C}^1(\Omega)$ and a minimizing region is a member of
${\cal C}^{2-}(\Omega)$.]\ \ Further, $\partial\hat{C}$ is $C^2$
everywhere except on the closed set of measure zero given by $\partial
{\cal W}$, where ${\cal W} = (\partial\Omega \cap \partial\hat{C})$
(and $\partial {\cal W}$ is constructed viewing ${\cal W}$ as a subset
of either $\partial\Omega$ or $\partial\hat{C}$).

Note that although $\partial {\cal W}$ is by its definition a closed
subset of $\partial\hat{C}$, its being a set of measure does not
appear to be guaranteed as there exist boundaries of positive measure.

In conjectures~1 and~2, we have asserted that the surface
$\partial\hat{C}$ is a $C^{2-}$ submanifold that is almost everywhere
$C^2$.  It is too much to expect that $\partial\hat{C}$ will be
everywhere $C^2$ as we expect a discontinuity in its mean extrinsic
curvature $H$ where $\partial\hat{C}$ ``first intersects''
$\partial\Omega$, i.e., on $\partial {\cal W}$.  To see this, suppose
conjectures~1 and~2 are true.  Then, write $\partial\hat{C}$ as the
disjoint union of three sets as follows
\begin{equation}
\partial\hat{C} = (\partial\hat{C} \setminus {\cal W})
\cup ({\cal W} \setminus \partial {\cal W})
\cup (\partial {\cal W}).
\end{equation}
As $\partial\hat{C} \setminus {\cal W}$ is in the interior of
$\Omega$, $H = \sigma(\hat{C})$ and $H = \kappa$ in conjectures~1
and~2, respectively.  However, as $\partial\hat{C}$ coincides with
$\partial\Omega$ on the open set ${\cal W} \setminus \partial {\cal
W}$, $H$ will equal the mean extrinsic curvature of $\partial\Omega$
on ${\cal W} \setminus \partial {\cal W}$.  Therefore, in general, we
expect that $H$ will suffer a discontinuity on $\partial {\cal W}$.
So, as $H$ will not always be $C^2$, $\partial\hat{C}$ will not always
be $C^2$.  However, note that this argument suggests that the lack in
continuity in the second-order partial derivatives defining the
surface arise from mere jumps and not divergences.  It is this
property that suggests that the surface is $C^{2-}$.

While we shall not attempt to do so here, conjectures~1 and~2 can
probably be proven using the ideas and techniques of geometric measure
theory \cite{gmt}.  Very roughly, we consider a subset ${\cal
V}(\Omega)$ of ${\cal C}^{1-}(\Omega)$ whose members are sufficiently
well-behaved that they have finite volume and surface area (using the
Hausdorff measure).  One then argues that $S$ is a continuous function
(in some natural topology) on ${\cal V}(\Omega)$ and that the subset
of ${\cal V}(\Omega)$ defined by those $C \in {\cal V}(\Omega)$ such
that $S(C) \le S(\Omega)$ is compact.  It then follows immediately
that there is a region $\hat{C} \in {\cal V}(\Omega)$ that achieves
the minimal value of $S$ on this set.  The last step would be to
establish that $\hat{C}$ is actually a member of ${\cal
C}^{2-}(\Omega)$ and $C^2$ on $\partial\hat{C} \setminus {\cal W}$
(and that ${\cal W}$ is a set of measure zero).  We leave the task of
showing that these steps can actually be completed open for
investigation.

\section{Future trapped regions are trapped}
\label{sec:trapped}

Although there exists theorems showing that future trapped regions
must lie within the black hole region of the spacetime, the arguments,
as given, require that their surfaces be everywhere $C^2$
\cite{Wald84,HawkingEllis73}.  Here, we show that the same result
holds for regions with boundaries that are not quite this smooth, and
so deserve to be called future trapped regions.  To make our method of
proof clear, we first cover the case where the surface of the region
is everywhere $C^2$.  After this, we modify the proof to accommodate
our more general regions.  We then discuss the possibility of further
generalizations.

While our method of proof is similar to the existing proofs for smooth
regions, there is a notable difference in the final derived
contradiction.  The Hawking and~Ellis argument ends with the
contradiction that the area of $\partial C$ is no less than the area
of $\partial J^+(C) \cap {\cal J}^+$, which, being at infinity, is
infinite.  The Wald argument ends with the contradiction that the
future expansion of the null generators of $\partial J^+(C)$ is
nonpositive on $\partial C$ and yet positive near ${\cal J}^+$.  Here,
we end with the contradiction that there are null generators of
$\partial J^+(C)$ extending beyond ${\cal J}^+$ that possess a point
conjugate to $\partial C$ on ${\cal J}^+$.

Actually, it should be noted that the Wald argument contains a slight
error in that the local cross-sections of ${\cal J}^+$ constructed
need not have the requisite differentiability properties in order that
nearby cross-sections of $\partial J^+(C)$ have strictly positive
future expansion.  A simple counterexample is provided by a smooth
closed region $C$ in a flat spatial hypersurface $\Sigma$ in Minkowski
spacetime with the property that all of $C$ lies to one side of a flat
plane ${\cal P}$ in $\Sigma$ except for a closed region $\partial C
\cap {\cal P}$ having a non-empty interior (as a subset of ${\cal
P}$).  Then, it is not difficult to see that the null generators of
$\partial J^+(C)$ having past endpoint on $\partial C \cap {\cal P}$
intersect ${\cal J}^+$ and have zero expansion everywhere.  Of course,
the Wald argument can easily be fixed by introducing an area type
argument, or by adopting the method of theorem~4, which can be viewed
as such a fix as it has much of its inspiration from the Wald
argument.

Our notion of asymptotic flatness is that given in Ref.~\cite{Wald84}.
We denote the manifolds of the ``physical'' and ``unphysical''
spacetime by $M$ and $M'$, respectively.  We remind the reader that $M
= M' \setminus (\overline{J^+(i^0)} \cup \overline{J^-(i^0)})$, where
$i^0$ is the point representing spatial infinity.  Therefore,
$\partial M = (i^0 \cup {\cal J}^+ \cup {\cal J}^-)$, where ${\cal
J}^{\pm} = (\partial J^{\pm}(i^0)) \setminus i^0$ are future and past
null infinity.

Furthermore, the theorems we prove are for strongly asymptotically
predictable spacetimes \cite{Wald84}, which are simply those
asymptotically flat spacetimes for which there exists an open globally
hyperbolic subset $V$ of $M'$ containing $\overline{J^-({\cal J}^+)
\cap M}$ (where the closure is as a subset of $M'$).  Note that
$\partial M \subset V$.  It can be shown that all globally hyperbolic
asymptotically flat spacetimes are strongly asymptotically
predictable.  Further, the globally hyperbolic asymptotic region $V$
can be chosen so that it contains all of $M$ and an asymptotically
flat Cauchy surface $\Sigma$ for $M$ together with spatial infinity
$i^0$ is a Cauchy surface for $V$.  Therefore, the requirement that a
subset $C$ of $\Sigma$ be closed and bounded away from infinity (so
there exists a neighborhood of $i^0$ disjoint from $C$) is equivalent
to the condition that $C$ be closed as a subset of $\Sigma' = (\Sigma
\cup i^0)$.

\subsection{Regions whose surfaces are $C^2$}
\label{sec:trapped_c2}

{\it Theorem 4.} Fix a smooth strongly asymptotically predictable
spacetime \cite{Wald84} satisfying the null-convergence condition
\cite{ncc}.  Let $\Sigma'$ be a smooth asymptotically flat Cauchy
surface for $V$ and let $C \subset (\Sigma' \cap M)$ be a future
trapped region in the sense that $C$ is a closed subset of $\Sigma'$,
$\partial C$ is $C^2$, and the convergence of the outward
future-directed null normals to $\partial C$ is everywhere
non-negative.  Then, $(C \cap J^-({\cal J}^+)) = \emptyset$.  [That
is, $C \subset (\Sigma' \cap B)$, where $B$ is the black-hole region
of the spacetime.]

{\it Proof.}  In the following, all of our constructions are carried
out solely within the asymptotic globally hyperbolic region $V$.
Therefore, statements regarding the openness or closedness of sets
refer to these properties in $V$ alone.  Since $C$ does not contain
$i^0$ (as $C$ is a subset of $M$), $J^+(C)$ does not contain $i^0$.
Further, $J^+(C)$ is closed, since $C$ is a closed subset of
$\Sigma'$.  (See exercise~8 from chapter~8 of Ref.~\cite{Wald84}.)\ \
Therefore, there is a neighborhood of $i^0$ disjoint from $J^+(C)$.

Suppose, for contradiction, that $(C \cap J^-({\cal J}^+)) \neq
\emptyset$.  Then, $(J^+(C) \cap {\cal J}^+) \neq \emptyset$, and,
hence, $(J^+(C) \cap I^+(i^0)) \neq \emptyset$.  It then follows that
$(\partial J^+(C) \cap I^+(i^0)) \neq \emptyset$.  To see this, fix
any point $p \in (J^+(C) \cap I^+(i^0))$.  Then, as there exists a
timelike curve $\gamma$ from $i^0$ to $p$ [which must lie entirely
within $I^+(i^0)$] and there exists an open neighborhood of $i^0$
disjoint from the closed set $J^+(C)$, the curve $\gamma$ must leave
$J^+(C)$ and therefore intersect $\partial J^+(C)$, showing that
$(\partial J^+(C) \cap I^+(i^0)) \neq \emptyset$.

Recall that if $p$ is any point on a null generator of $\partial
J^+(C)$ whose past endpoint on $\partial C$ has an open neighborhood
on which $\partial C$ is $C^2$, there must not be a point conjugate to
$\partial C$ between $\partial C$ and $p$
\cite{Wald84,HawkingEllis73}.  Pick a point $p \in (\partial J^+(C)
\cap I^+(i^0))$ and a null generator $\nu$ of $\partial J^+(C)$
containing $p$.  Then $\nu$ cannot possess a point conjugate to
$\partial C$ in $M$ (with respect to either the physical or unphysical
metric) nor on ${\cal J}^+$ (with respect to the unphysical metric).
However, in the physical portion of the spacetime $M$, it follows from
the null Raychaudhuri equation and the null-convergence condition that
the (physical) future convergence of the null generators of $\partial
J^+(C)$ is not only non-negative on $\partial C$, it is non-negative
everywhere to the future \cite{Wald84,HawkingEllis73}.  Furthermore,
if such a generator has positive convergence $\rho_0 > 0$ at some
point, then it must possess a conjugate point within an affine
parameter time $2/\rho_0$ thereafter, provided the generator can be
extended this far.  Therefore, as $\nu$ is future complete in the
physical metric in $M$ (as it intersects ${\cal J}^+$), the (physical)
convergence along $\nu$ must be zero in $M$.  Therefore, in the
infinitesimal sense, the physical area of a bundle of outgoing
future-directed null rays orthogonal to $\partial C$ is constant along
$\nu$ (in $M$).  In terms of the unphysical metric, this area is that
given by the physical area multiplied by the square of the conformal
factor.  As this conformal factor is zero on ${\cal J}^+$, it follows
that $\nu$ possesses a point conjugate to $\partial C$ where it
intersects ${\cal J}^+$ (with respect to the unphysical metric), which
is a contradiction.~$\Box$

\subsection{Regions whose surfaces are not quite $C^2$}
\label{sec:trapped_notc2}

The problem with the proof of theorem~4 when $\partial C$ is not
everywhere $C^2$ is that it may happen that because of our choice of
$p$ in the last paragraph, $\nu$ may have its past endpoint at a place
on $\partial C$ where the surface is not $C^2$, thus making the final
conjugate point argument inapplicable.  When $\partial C$ is
everywhere $C^{2-}$ and $C^2$ on $\partial C \setminus {\cal Z}$,
where ${\cal Z}$ is a closed set of measure zero, although we do not
have complete freedom in what choice to make for $p$, it turns out we
can always find one so that the past endpoint of its associated null
generator has a neighborhood within $\partial C$ on which the surface
is $C^2$, i.e., its past endpoint is somewhere on $\partial C
\setminus {\cal Z}$.  The idea is that it is impossible for only the
generators of $\partial J^+(C)$ with past endpoint on ${\cal Z}$ to
make it beyond ${\cal J}^+$ as there are not ``enough of them'' to
make up a ``local piece'' of $\partial J^+(C)$, as ${\cal Z}$ is a set
of measure zero in $\partial C$.

We capture this idea using the notion of Hausdorff measure \cite{gmt}.
On a differentiable manifold $N$ with Riemannian metric, for any two
points $a$ and $b$ in $N$, define $d(a,b)$ to be the greatest lower
bound on the lengths of $C^1$ curves in $N$ connecting $a$ to $b$ [so
$(N,d)$ is a metric space].  For any subset $S \subset N$, set
$\text{diam}(S) = \sup_{a,b \in S}(d(a,b))$.  Then, for any subset $A
\subset N$ and numbers $k$ and $\delta > 0$, set
\begin{equation}
{\cal H}^k_\delta(A) = \inf \sum_j \nu_k
\left( {\text{diam}(S_j) \over 2} \right)^k,
\end{equation}
where $\nu_k$ is the volume of a unit-ball in flat ${\Bbb R}^k$ when
$k$ is a non-negative integer (so $\nu_0 = 1$, $\nu_1 = 2$, $\nu_2 =
\pi$, $\nu_3 = 4\pi/3$, etc.) and an arbitrary positive constant
otherwise, and where the infimum is taken over over all countable
coverings $\{ S_j \}$ of $A$ (i.e., $A \subset \cup_j S_j$) with
$\text{diam}(S_j) \le \delta$.  With this, the ${\cal H}^k$-measure of a
set $A$ is defined as
\begin{equation}
{\cal H}^k(A) = \lim_{\delta \rightarrow 0} {\cal H}^k_\delta(A).
\end{equation}
This limit is well defined (though possibly infinite) as ${\cal
H}^k_\delta(A)$ is non-decreasing in $\delta$.  It is worth noting
that if ${\cal H}^k(A) < \infty$ then ${\cal H}^m(A) = 0$ for all $m >
k$.  It can be shown that if $A$ is an $k$-dimensional $C^1$ embedded
submanifold of $N$ with $k \le \dim(N)$, then ${\cal H}^k(A)$
corresponds to the usual ``volume'' of this submanifold.  For
instance, in the case $\dim(N)=3$, ${\cal H}^1(A)$ is the length of a
1-dimensional submanifold $A$, ${\cal H}^2(A)$ is the area of a
2-dimensional submanifold $A$, and ${\cal H}^3(A)$ is the volume of a
3-dimensional submanifold $A$.

With this, we say a subset $A$ of a differentiable manifold $N$ has
${\cal H}^k$-measure zero if ${\cal H}^k(A) = 0$.  It can be shown
that this notion is independent of which Riemannian metric is chosen,
and, therefore, whether a subset of a (paracompact) manifold has
${\cal H}^k$-measure zero is dependent solely upon the set.  In the
case where $k = \dim(N)$, ${\cal H}^k$-measure zero is identical to
the usual Lebesgue notion of measure zero on a differential manifold.
Furthermore, if $f$ is a locally Lipschitz map from the manifold $N$
to another differentiable manifold, it follows that if ${\cal H}^k(A)
= 0$, then ${\cal H}^k(f[A]) = 0$.

Using these concepts, we can now prove that our generalized future
trapped regions are indeed trapped.

{\it Theorem 5.} Fix a smooth strongly asymptotically predictable
spacetime \cite{Wald84} satisfying the null-convergence condition
\cite{ncc}.  Let $\Sigma'$ be a smooth asymptotically flat Cauchy
surface for $V$ and let $C \subset (\Sigma' \cap M)$ be a future
trapped region in the sense that $C$ is a closed subset of $\Sigma'$,
$\partial C$ is everywhere $C^{2-}$ and, on $\partial C \setminus {\cal
Z}$, $\partial C$ is $C^2$ and the convergence of the outward
future-directed null normals to $\partial C$ is non-negative, where
${\cal Z}$ is a closed set of measure zero.  Then, $(C \cap J^-({\cal
J}^+)) = \emptyset$.  [That is, $C \subset (\Sigma' \cap B)$, where
$B$ is the black-hole region of the spacetime.]

{\it Proof.}  Suppose, for contradiction, that $(C \cap J^-({\cal
J}^+)) \neq \emptyset$.  Then, using the same argument as in
theorem~4, it again follows that $(\partial J^+(C) \cap I^+(i^0)) \neq
\emptyset$.  We claim that there exists $p \in (\partial J^+(C) \cap
I^+(i^0))$ with an associated null generator $\nu$ having past
endpoint on $\partial C \setminus {\cal Z}$, an open subset of
$\partial C$ where the surface is $C^2$.  We show this by arguing that
there are not enough generators with past endpoint on ${\cal Z}$ to
make up $\partial J^+(C)$ in $I^+(i^0)$ as follows.

First, the subset $\tilde{\cal Z}$ of $\partial J^+(C)$ consisting of
those points with null generators having past endpoint on ${\cal Z}$
has ${\cal H}^3$-measure zero.  To see this, denote by ${\cal K}$ the
subset of $TV$ (the tangent bundle associated with $V$) consisting of
all pairs $(p,k^a)$ where $p \in \partial C$ and $k^a$ is an outward
future-directed null vector normal to $\partial C$ at $p$.  Using the
fact that $\partial C$ is $C^{2-}$, it follows that there exists a
locally Lipschitz map from $\partial C \times {\Bbb R}$ onto ${\cal K}
\subset TV$.  Next, since $\partial J^+(C) \setminus C$ is generated
by null geodesics with past endpoint on $\partial C$ and
future-directed outgoing tangent vector normal to $\partial C$, we see
that $\partial J^+(C) \setminus C$ is a subset of the projection of
$\exp({\cal K})$ onto $V$ (where $\exp$ is the smooth diffeomorphism
from $TV$ to $TV$ defined by the geodesic flow on $TV$).  As both
$\exp$ and the projection map are smooth, it follows that $\partial
J^+(C) \setminus C$ is a subset of the image of a subset of $\partial
C \times {\Bbb R}$ under a locally Lipschitz map.  Therefore, since
${\cal Z} \times {\Bbb R}$ has ${\cal H}^3$-measure zero as a subset
of $\partial C \times {\Bbb R}$ (which follows from the fact that
${\cal Z}$ has ${\cal H}^2$-measure zero as a subset of $\partial C$)
and since $\tilde{\cal Z}$ is a subset of the image of a subset of
${\cal Z} \times {\Bbb R}$ under a locally Lipschitz map, it follows
that $\tilde{\cal Z}$ has ${\cal H}^3$-measure zero in $V$.  (Note
that it is in the establishment of this result that we use the fact
$\partial C$ is $C^{2-}$ and not merely $C^1$.)

Next, pick any point $q \in (\partial J^+(C) \cap I^+(i^0))$ and an
open neighborhood $O$ of $q$ with $O \subset I^+(i^0)$.  Using the
fact that $\partial J^+(C)$ is an achronal $C^{1-}$ embedded
three-dimensional submanifold of $V$ (see proposition~6.3.1 of
Ref.~\cite{HawkingEllis73}), it follows that $\partial J^+(C) \cap O$
has positive ${\cal H}^3$-measure.  (To see this, note that we can
choose $O$ so that it is diffeomorphic to an open subset of ${\Bbb
R}^4$ with $\partial J^+(C) \cap O$ corresponding to the graph of a
$C^{1-}$ function of three variables.)\ \ Therefore, as the subset of
$\partial J^+(C)$ consisting of generators with past endpoint on
${\cal Z}$ has ${\cal H}^3$-measure zero, it follows that there must
exist a point $p \in \partial J^+(C) \cap O$ with an associated null
generator $\nu$ that has past endpoint on $\partial C \setminus {\cal
Z}$.  (In fact, there are many such points.)

Arguing as we did in theorem~4 shows that $\nu$ contains a point
conjugate to $\partial C$ (with respect to the unphysical metric)
where $\nu$ intersects ${\cal J}^+$ (being between $\partial C$ and
$p$), which is a contradiction.~$\Box$

\subsection{Possible generalizations}
\label{sec:generalize}

In extending the notion of a future trapped region, we have restricted
ourselves to regions $C$ with $C^{2-}$ surfaces that are further $C^2$
everywhere except on a closed set of measure zero.  We have done this
because this is both what we expect of the surfaces constructed
(conjectures~1 and~2) and these are regions for which we can carry
through all the relevant arguments (theorems~3 and~5).  However, a
much greater extension seems possible.  For instance, it is plausible
that the notion of a future trapped region can be extended to regions
with surfaces that are merely $C^{2-}$.  Such a surface is twice
differentiable everywhere except on a set of measure zero ${\cal Z}$.
If the convergence of a family of future-directed outgoing null
geodesics orthogonal to a surface can be defined on $\partial C
\setminus {\cal Z}$ and the conjugate point argument used in theorem~5
can be applied to the generators with past endpoint on $\partial C
\setminus {\cal Z}$, the notion of a future trapped region with a
$C^{2-}$ surface would be a well-defined concept.

However, it would seem that the best notion of a region being future
trapped would not involve any differentiability conditions.  For
example, consider the analogous problem of what we mean by a closed
region $C$ in flat space having a surface $S$ that is everywhere
locally convex.  Here, we have a precise notion that imposes no
differentiability conditions on the surface: For each point $p \in S$
there is a neighborhood $N$ in $S$ such that $(1-\lambda) x + \lambda
y \in C$ for all $x,y \in N$ and $\lambda \in [0,1]$.  (That is, the
convex hull of $N$ is a subset of $C$.)\ \ Likewise, we say the
surface of a region $C$ is locally concave if it is locally convex
when viewed as the surface of the closure of the complement of $C$.
Note that this flat space notion has a natural generalization to
curved spaces: We call the surface $S$ of a closed region $C$ locally
convex if for each point $p \in S$ there is a neighborhood $N$ in $S$
and a convex normal neighborhood $U$ containing $N$ such that for all
points $x, y \in N$ the geodesic from $x$ to $y$ (within $U$, being
unique) lies within $C$.  In the $C^2$ case, the above implies the the
extrinsic curvature $H_{ab}$ of $S$ is positive semi-definite.

We want a geometric condition that, in the $C^2$ case, leads to the
bound $H = H^a{}_a \le K_{ab}h^{ab}$.  Surely, such a notion would be
based on a demand that the areas of all local cross sections of
$\partial J^+(C)$ are non-increasing to the future (at least
sufficiently near $\partial C$).  The problem is to capture this idea
in a well-defined sense.  For instance, one needs for $\partial
J^+(C)$ to be sufficiently well-behaved so that the surface areas of
suitable cross-sections are well-defined.  This is probably not such a
problem as $\partial J^+(C)$ is an imbedded $C^{1-}$ submanifold for
any set $C$.  Then, to show that such regions are indeed trapped, an
area-type argument similar to that used by Hawking and~Ellis would
probably be the most natural method to use.  However, how is one to
show that the areas of cross-sections are non-increasing to the future
when the null Raychaudhuri equation cannot be implemented?  Clearly,
some subtlety is needed here.

Note that a naive condition such as $\partial C$ being everywhere
$C^{1-}$ and, on $\partial C \setminus {\cal Z}$, $\partial C$ is $C^2$
and the convergence of the outward future-directed null normals to
$\partial C$ is non-negative, where ${\cal Z}$ is a closed set of
measure zero, is insufficient.  A simple counterexample is provided by
taking $C$ to be a solid cube in a flat spatial hypersurface in
Minkowski spacetime.  Here, $\partial C$ is everywhere $C^{1-}$ and,
except along the edges and vertices (a closed set ${\cal Z}$ of
measure zero), the surface is $C^\infty$ and the convergence of the
outward future-directed null normals to $\partial C$ is zero.
However, $C$ is clearly ``visible'' from ${\cal J}^+$, i.e., it is not
trapped.  In the proof of theorem~5, the problem with such surfaces is
that one does not have a one-to-one correspondence between the null
generators of $\partial J^+(C)$ and $\partial C$, and, as a result,
the portion of $\partial J^+(C)$ consisting of the generators having
past endpoint on ${\cal Z}$ has positive ${\cal H}^3$-measure.  For
example, at a vertex, an entire ``octant's worth'' of null generators
of $\partial J^+(C)$ intersect $\partial C$ at a single point.  In
this case, all null generators of $\partial J^+(C)$ that do make it
beyond ${\cal J}^+$ have past endpoints on ${\cal Z}$.

Lastly, one might expect that a differentiability condition that would
be sufficient to establish that a region $C$ is future trapped is that
$\partial C$ is everywhere $C^1$ and $C^2$ on an open dense subset $D$
of $\partial C$ (with the convergence of the outward future-directed
null normals being non-negative on $D$).  In fact, this was the
approach first taken herein, but was abandoned because of a
difficulty.  The idea is that if a null generator $\nu$ associated
with a point $p \in (J^+(C) \cap I^+(i^0))$ has its past endpoint on
$D$, the argument proceeds as in theorem~4, while if not, then it
would seem that we could find a point arbitrarily near $p$ in $(J^+(C)
\cap I^+(i^0))$ with an associated generator having past endpoint on
$D$.  (After all, $D$ is dense in $\partial C$.)\ \ While this may be
true, proving it appears to be difficult.  For instance, although one
might expect that there would exist a neighborhood of $\nu \cap
\partial C$ (within $\partial C$) such that all null generators with
past endpoint thereon remain on $\partial J^+(C)$ long enough to enter
$I^+(i^0)$, it turns out that this need not be the case if we just use
the fact that $\partial C$ is $C^1$.  Whether this does hold when the
additional conditions on $\partial C$ are used is not clear.

\section{Discussion} \label{sec:discussion}

Theorems~1 and~3 provide us with simple tests for the existence of
future trapped regions within an initial data set, but how effective
are they?

First, the conditions of theorems~1 and~3 are quite strong in the
following sense.  Recall that theorem~1 requires that
$\min_\Omega(\kappa) \ge \sigma(\Omega)$ (as does Eardley's theorem).
Using the fact that $\kappa$ is the sum of the two lesser principal
(extrinsic) curvatures $(k_1,k_2,k_3)$, it is not difficult to show
that $K^a{}_a = k_1+k_2+k_3 \ge {3 \over 2} \kappa \ge {3 \over 2}
\sigma(\Omega) > 0$ everywhere on $\Omega$, showing that this region
is everywhere contracting ``on average''.  However, if $K^a{}_a$ is
non-negative, $\kappa$ need not be positive.  This shows that the
region $\Omega$ is more than contracting ``on average''.  Indeed, on a
maximal hypersurface ($K^a{}_a = 0$ everywhere on $\Sigma$), $\kappa
\le 0$ (with equality only where $K_{ab}=0$) everywhere on $\Sigma$.
In this respect, the condition of theorem~1 (and Eardley's theorem) is
quite strong.  While theorem~3 merely requires that $S(\Omega) \le 0$,
so $\kappa$ need not be positive on all of $\Omega$, $\kappa$ still
must be positive over a sufficiently large subset of $\Omega$ in order
to meet this condition.

Second, while both theorems give sufficient conditions for the
existence of future trapped regions, neither condition is necessary.
This is easily seen by constructing a momentarily static initial data
set (so $K_{ab} = 0$, and hence $\kappa = 0$, on $\Sigma$) that
contains a minimal two-surface bounding a compact region $C$.  This
region $C$ is future (and past) trapped and yet, as $S(C) = A(C)$ is
positive, the condition of neither theorem~1 nor~3 is met.

Third, neither theorem is very sensitive to the ``local''
existence of a future trapped region in the following sense.  Suppose
we have a future trapped surface $S$ such that both families of
future-directed orthogonal null congruences have strictly positive
convergence on $S$.  Construct a three-dimensional region $\Omega$ by
``thickening'' $S$ a small distance $r$ within an initial data surface
containing $S$.  Then, for $r$ sufficiently small, $\Omega$ is a
future trapped region.  However, for sufficiently small $r$,
$\sigma(\Omega)$ will be larger than $\inf_\Omega(\kappa)$ and
$S(\Omega)$ will be positive, and hence neither theorem enables us to
deduce that $\Omega$ itself is a future trapped region.

Fourth, the conditions of theorems~1 and~3 are quite robust in the
sense that if we have a region $\Omega$ that satisfies the condition
of either theorem with strict inequality and then deform it to create
a new region $\Omega'$ by ``pushing'' very thin fingers of the surface
of $\Omega$ into $\Omega$ (in arbitrarily complex ways), then,
provided our fingers are sufficiently thin, the surface area, volume,
and the integral of $\kappa$ for $\Omega'$ will be sufficiently near
those of $\Omega$ so that the conditions of both theorems will be met
for $\Omega'$.  More generally, if we construct $\Omega'$ by excising
sufficiently thin regions from $\Omega$, both theorems guarantee the
existence of a future trapped region within $\Omega'$.  This is
perhaps somewhat surprising at first given that $\Omega'$ can be
topologically quite complex. However, noting that the mean curvature
of the portions of the surface of $\Omega'$ created by excising ``very
thin fingers'' is very large and negative, we realize that $\hat{C}$
less the thin regions is nearly a future trapped region---all that is
needed is a bit of adjusting near the edges where the excised region
intersects $\hat{C}$.

Fifth, and last, theorems~1 and~3 do have a slight advantage in
numerical search for the existence of future trapped regions as
the calculation of $\sigma(\Omega)$ or $S(\Omega)$ requires only the
calculation of a surface area and a volume integral, which are not as
sensitive to numerical inaccuracies that would arise in calculating
the mean extrinsic curvature $H$ of a two-surface in $\Sigma$ to test
whether the surface is future outer trapped (i.e., testing whether the
condition given by Eq.~(\ref{ftr}) holds on the boundary).

So, while theorems~1 and~3 do offer tests for the existence of future
trapped regions, their inability to detect the existence of future
trapped regions in some instances, e.g., in initial data sets
associated with maximal hypersurfaces and ``thin'' future trapped
regions, leads us to wonder whether stronger tests of the type
considered here can be devised to give sufficient conditions for the
existence of future trapped regions.

\acknowledgements I thank Douglas Eardley, for his encouragement in
this work and for allowing me to visit the ITP at UCSB for its
discussion, and Robert Geroch, for answering numerous questions.


\begin{references}
\bibitem{ids}
Here, $\Sigma$ is an orientable three-manifold (without boundary),
representing the Universe at a given time; $g_{ab}$ is a Riemannian
metric on $\Sigma$, giving the intrinsic geometry of $\Sigma$ as
induced by the spacetime metric; and $K_{ab}$ is a symmetric tensor,
giving the extrinsic curvature of $\Sigma$ in the spacetime.
\bibitem{ncc}
A spacetime with Ricci tensor $R_{ab}$ is said to satisfy the
null-convergence condition if $R_{ab}k^ak^b \ge 0$ for all null $k^a$.
\bibitem{mean} Actually, this is a slight abuse in terminology in that
$H$ is the trace of extrinsic curvature of the two-surface in the
three-manifold and not one-half this value. (The latter quantity is
the mean value of the two principal curvatures.)
\bibitem{apparent}
This assumes that the future apparent horizon ${\cal A}$ has the
structure of a sufficiently smooth manifold.  See, e.g.,
Proposition~12.2.5 of Ref.~\cite{Wald84}.
\bibitem{projop} We call a tensor $h^a{}_b$, acting on a vector space
$V$ with metric $g_{ab}$, a rank~2 orthogonal projection map if
$h_{ab}$ is symmetric, $h^a{}_b = h^a{}_c h^c{}_b$, and $h^a{}_a = 2$,
where indices are raised and lowered with $g_{ab}$.  The trace
condition on $h^a{}_b$ ensures that $h^a{}_b$ maps $V$ onto a
two-dimensional vector subspace of itself, i.e., its rank is two.  It
is not difficult to show that with $\dim(V)=3$, there exists a unique
(up to sign) unit vector $n^a$ such that $h^{ab} = g^{ab} - n^a n^b$.
Further, any such unit vector $n^a$ defines a unique rank~2 orthogonal
projection map in this way.  While perhaps more complicated than it
could be otherwise, this statement of the theorem follows the spirit
of its proof.
\end{references}
\end{document}